\documentstyle[twocolumn,aps]{revtex}
\input psfig
\begin{document}
\newcommand {\be}{\begin{equation}}
\newcommand {\ee}{\end{equation}}
\newcommand {\bea}{\begin{eqnarray}}
\newcommand {\eea}{\end{eqnarray}}
\newcommand {\nn}{\nonumber}

\draft
\twocolumn[\hsize\textwidth\columnwidth\hsize\csname @twocolumnfalse\endcsname
%
%
%

\title{ Extended Gapless Regions in Disordered
${\rm d_{x^2-y^2}}$-Wave Superconductors}

\author{ Stephan Haas$^{(1)}$, A.V. Balatsky$^{(2)}$,
Manfred Sigrist$^{(1)}$, and T.M. Rice$^{(1)}$}

\address{$^{(1)}$ Theoretische Physik,
ETH-H\"onggerberg, CH-8093 Z\"urich, Switzerland,\\
$^{(2)}$ Theory Division, Los Alamos National Laboratory, Los Alamos, 
NM 87545, USA}

\date{\today}
\maketitle

\begin{abstract}

A generalization of the Abrikosov-Gorkov equations for non-magnetic
impurities in unconventional superconductors is proposed, including higher
harmonics in the expansion of the momentum dependent gap function and
a momentum dependent impurity scattering potential.
This model is treated within a self-consistent calculation to obtain the
electronic density of states, the optical conductivity,
and the gap function in a two-dimensional
${\rm d_{x^2-y^2}}$-wave
superconductor.
It is argued that momentum dependent scattering
from the impurities may lead to extended gapless
regions in the gap function centered around the nodes of the pure
${\rm d_{x^2-y^2}}$-wave superconductor.
The associated enhancement of the residual density of states may be
responsible for the rapid decrease of $\rm T_c$ and the increase
of the London penetration depth with hole doping observed
in overdoped
cuprate superconductors.

\end{abstract}

\pacs{}
\vskip2pc]
\narrowtext

The pair-breaking role of impurities in d-wave superconductors in the
weak coupling limit is well known. Recently the evidence that such weak
coupling descriptions apply in the overdoped region of the high-$\rm T_c$
cuprates has been growing.\cite{uchida}
However, the experiments that track the behavior
of the gap function have come to divergent conclusions.\cite{uchida,arpes}
This raises the question of the evolution of the gap function in a d-wave
superconductor as a function of the pairing interaction strength and the
impurity concentration.
In this letter we examine this question,
allowing for angular dependent impurity scattering potentials. Our conclusion 
is that the evolution of the gap function depends on the character
of the impurity scattering potential. In particular, we find that extended
gapless regions grow as the critical point is reached when the impurity 
scattering potential is weighted towards forward scattering. In this case
the scattering processes near the gap nodes, which connect regions of 
opposite sign pairing amplitude, are of increasing importance and reduce the
gap in this region. This contrasts with the regions around the gap maxima,
where the impurity scattering is less effective. The resulting evolution of
the gap function is sketched in Fig. 1.

In conventional Abrikosov-Gorkov (AG) theory, 
the interaction of electrons with non-magnetic 
impurities is governed by a momentum independent scattering
potential.\cite{ag}
Here we generalize this approach to the momentum dependent
case, with a scattering potential ${\rm u_{{\bf k},{\bf k'}} }$. 
The contribution to the self-energy
of electrons scattering
from this static potential is given by
the self-consistent T-matrix equation, 
\bea
{\rm \hat{T}_{{\bf k},{\bf k'}}(i\omega_n) = 
u_{{\bf k},{\bf k'}} \hat{\sigma}_3 + 
\sum_q u_{{\bf k},{\bf q}}
\hat{\sigma}_3 \hat{G}_{\bf q}(i\omega_n) \hat{T}_{{\bf q},
{\bf k'}}(i\omega_n), }
\eea
where ${\rm \hat{G}_{\bf k}(i\omega_n) }$ is the Nambu
single particle Green's function
in a superconductor,
\bea
{\rm
\hat{{G}}_{\bf k}(i\omega_n) =  \frac{{i\omega_n}\hat{\sigma}_0
+ {\Delta}_{\bf k}\hat{\sigma}_1 + {\xi}_{\bf k}
\hat{\sigma}_3 }
{ (i{\omega_n})^2 - {\Delta}_{\bf k}^2 - {\xi}_{\bf k}^2} ,
}
\eea
and the Pauli matrices ${\rm \hat{\sigma}_i }$ (${\rm \hat{\sigma}_0
= {\hat{\bf 1}} }$) form a complete basis in Nambu space.
The renormalization of the bare Green's function is given by
the self energy
${\rm \hat{\Sigma}_{\bf k} (i\omega_n) = \Gamma 
\hat{T}_{{\bf k},{\bf k}}(i\omega_n) }$, where ${\rm 
\Gamma }$ is the concentration of impurities. In the limits
${\rm |u_{\bf k,k'}| \rightarrow 0}$ (Born limit) 
and ${\rm |u_{\bf k,k'}| \rightarrow \infty}$ (unitary limit)
the T-matrix is simplified considerably. Hence, these 
limiting cases have been
studied extensively in the literature.\cite{sun,hotta}
However, 
the strength of the scattering potential is not known from first
principles, and only comparison
with experiments indicates that some impurities in high-${\rm T_c}$
cuprates, 
such as Zn,
are closer to the unitary limit.\cite{hirschfeld}

\begin{figure}[hp]
\centerline{\psfig{figure=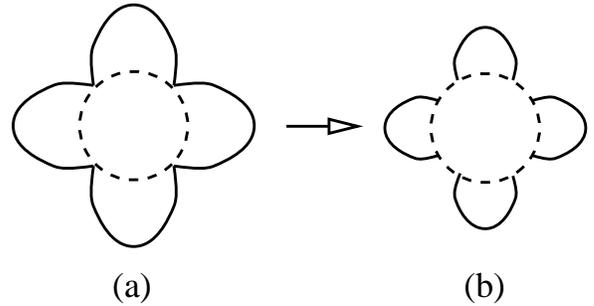,height=4cm,angle=-90}}
\vspace{0.5cm}
\caption{
Schematic illustration of the momentum dependence of the gap function in a 
${\rm d_{x^2-y^2}}$ superconductor (a) without disorder ($\Gamma =0$), and
(b)  in the presence of non-magnetic impurities ($\Gamma \ne 0$).}
\end{figure}

For computational simplicity, the Fermi surface is assumed
to have a cylindrical shape as shown in Fig. 1, neglecting effects of
the
more complicated material-dependent band structure observed
in high-${\rm T_c}$ compounds.\cite{pickett}
In this case the wave vector {\bf k} at the Fermi surface can be simply
expressed as an azimuthal angle, $\phi $.
This simplifying
assumption of a large and approximately cylindrical Fermi surface is
in qualitative agreement with photoemission and Hall angle experiments on
overdoped and even optimally doped cuprate samples.\cite{arpes,nishikawa}

The equations for the single particle Green's function,
${\rm {\hat{\tilde{G}}}}$${\rm _{\bf k}
^{-1}(i\omega_n) = \hat{G}^{-1}_{\bf k}(i\omega_n) 
- \hat{\Sigma}_{\bf k}(i\omega_n) }$,
have to be solved  simultaneously with the
zero-temperature gap equation, ${\rm \Delta_{\bf k '} =
2 \int_0^{\omega_D} d\omega_n \sum_{\bf k}
V_{\bf k,k'} \tilde{\Delta}_{\bf k}/
\sqrt{\tilde{\omega}_n^2 + \tilde{\Delta}_{\bf k}^2} }$,
to yield both the amplitude of the gap function and the 
components of the renormalized Green's function
at a given
impurity concentration ${\rm \Gamma }$.\cite{footnote1}
Here, ${\rm V_{\bf k,k'} }$ is the effective pairing 
potential in the d-wave channel
of the same functional form
as ${\rm \Delta_{\bf k} }$.
To describe the angular evolution of the gap function, we allow for
a higher harmonic, $\Delta_{\bf k } =$ $\Delta_0 \cos(2 \phi ) +$
$\Delta_1 \cos(6 \phi ) $, and simultaneously include such
higher harmonics in $\rm V_{\bf k,k'} $, which we parametrize in
factorized
form, 
${\rm V_{\bf k,k'} =
V(\cos(2\phi) + \alpha \cos(6\phi)) }$${\rm
(\cos(2\phi') + \alpha \cos(6\phi')) }$, where $\alpha$ is the parameter
which determines the gap shape in the absence of impurities.

In the following, we analyze how the renormalization due to
electron-impurity scattering affects the amplitude and the 
in particular the {\it shape}
of the gap function, $\rm \Delta_{\phi} = $ $\rm \Delta_0 \cos(2 \phi ) +$
$\rm \Delta_1 \cos(6 \phi )$. Let us examine the Born limit. 
For the choice of the impurity scattering potential,
${\rm \hat{u}_{\bf k,k'} = \hat{\sigma}_3
[u_0 + u_1cos(\phi - \phi ')] }$,
the AG equations take the form
\bea
{\rm \tilde{\omega}_n = \omega_n + \Gamma \langle
\frac{\tilde{\omega}_n (u_0^2 + u_1^2 cos^2 \phi ) }
{\sqrt{\tilde{\omega}_n^2 + \tilde{\Delta}_{\phi}^2}  }
\rangle }, \\
{\rm \tilde{\Delta}_{\phi'} = \Delta_{\phi'} - \Gamma \langle
\frac{\tilde{\Delta}_{\phi} u_1^2 cos(2\phi) }
{2 \sqrt{\tilde{\omega}_n^2 + \tilde{\Delta}_{\phi}^2}  }
\rangle \cos(2 \phi ') }.
\eea 
Here $\langle ... \rangle$ denotes the angular average over the Fermi 
surface, and constants arising from the radial momentum integration
are absorbed in $\Gamma$. Note that $\rm \tilde{\Delta}_{\phi} $ depends
on $\rm \tilde{\omega}_n$ (r.h.s. of Eq. (4)), and hence the 
actual shape of the
renormalized gap can be defined only by analyzing the 
corresponding angular dependent density of states (DOS).

There is a gap equation for each of the two components of
$\rm \tilde{\Delta}_{\phi} $. However, due to our particular choice of
the impurity scattering potential only $\rm \Delta_0$ is renormalized,
\bea 
\rm \Delta_0 = 2 N_0 V \int_0^{\omega_D} d\omega_n
\langle \frac{(\cos(2\phi) + \alpha \cos(6\phi)) \tilde{\Delta}_{\phi} }
{\sqrt{\tilde{\omega}_n^2 + \tilde{\Delta}_{\phi}^2}  }
\rangle ,
\eea
while the gap equation for the second harmonic gives
$\rm \Delta_1 = \alpha \Delta_0$. Thus the renormalized gap function 
is of the form: $\rm \tilde{\Delta}_{\phi} (i\omega_n) =$ $\rm 
\tilde{\Delta}_0 (i\omega_n) 
\cos(2\phi) + $ $\rm \alpha \Delta_0 \cos(6\phi) $. Since only the first 
component of the gap function is reduced by impurity scattering,
Eq. (4), it is
evident that the node region widens with increasing
impurity concentration ($\rm \Delta_1 / \Delta_0 \rightarrow 1/4$). 

\begin{figure}[hp]
\centerline{\psfig{figure=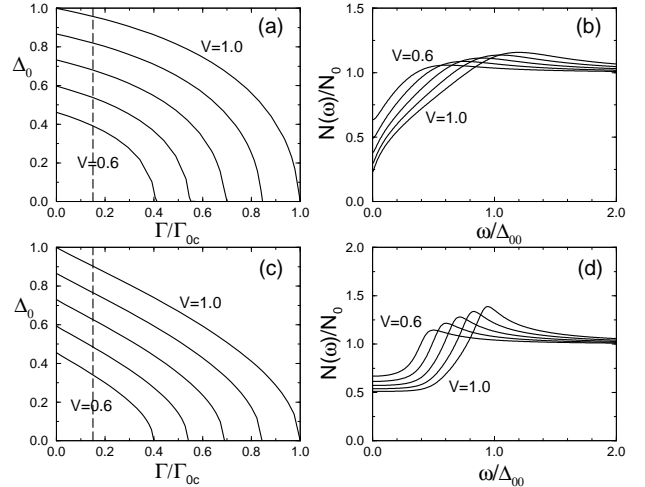,height=6.5cm,angle=-90}}
\vspace{0.5cm}
\caption{
(a) Amplitude of the gap function in a disordered ${\rm d_{x^2-y^2}}$-wave
superconductor as a function of the impurity concentration in the Born limit.
A pure forward-scattering impurity potential is chosen
($\rm u_0 = u_1 =1$), and the gap anisotropy parameter is set to 
$\rm \alpha = 0.2$. The gap amplitude is plotted relative to its value
in the absence of impurities,
$\rm \Gamma =0$, and V=1.0. The impurity concentrations are measured with 
respect 
to $\rm \Gamma_{0c}$,
the critical concentration at V=1.0.
(b) DOS corresponding to the specific impurity concentration
$\rm \Gamma / \Gamma_{0c} =0.15$, indicated by the dashed line in (a).
$\rm N_0$ is the DOS in the normal state, and
$\rm \Delta_{00} \equiv \Delta (\Gamma = 0, V =1.0)$
(c) As (a), but for unitary impurity scattering with
$\rm u_0 = 1$, $\rm u_1 = 0$, and $\alpha =0$. 
(d) As (b), but for the parameters used in (c).
}
\end{figure}

In Fig. 2(a) the gap amplitude $\Delta_0$ obtained from the
numerical solution of Eqs. (3-5) is shown as a function of impurity
concentration and of the strength of the 
pairing interaction.
Here we have chosen $\alpha$=0.2, $\rm u_0$ = 1, and
$\rm u_1$ = 1, i.e. the case favoring forward scattering. 
The superconducting gap can be destroyed both by increasing the impurity
concentration
beyond a critical value, $\rm \Gamma_c $, or by reducing the pairing     
potential V below a critical $\rm V_c $.
While the first case naturally occurs when introducing disorder
into samples by impurity doping, the latter case may be realized
by introducing additional holes into overdoped cuprate
superconductors.\cite{footnote2} 
 
From an expansion of Eqs. (3-5) about $\rm \Gamma_c$ 
we find that for large impurity concentrations the 
amplitude of the gap function vanishes as $\rm \Delta_0 \propto 
\sqrt{1 - \Gamma / \Gamma_c } $. Similarly, if one drives the system 
critical by reducing $\rm V/V_c $, we find
$\rm \Delta_0 \propto \sqrt{V/V_c - 1} $, in agreement with the
full numerical solution of these equations.

From an analytic continuation of Eqs. (3) and (4) onto the real frequency
axis, the electronic DOS can be evaluated as
$\rm N(\omega ) = -\frac{1}{\pi }\sum_{\bf k}Im[ 
{\hat{\tilde{G}}}_{\bf k}
(i\tilde{\omega}_n) ]|_{i\tilde{\omega}_n  =
\tilde{\omega} + i\delta} $. In Fig. 2(b) the DOS is shown at 
$\rm \Gamma /\Gamma_c$ = 0.15 (indicated by the dashed line in Fig. 2(a) ).
As the strength of the pairing potential
is lowered, the system approaches the critical region where 
the gap is small, and hence the residual DOS increases with
decreasing V. As will be discussed in the following, this increase in
N(0) is enhanced by the occurrence of flat regions around
the nodes.

The above results were obtained in the Born limit. In the unitary limit,
the AG equations do not
reduce to simple expressions like
Eqs. (3-5) if the impurity scattering potential is taken to be
momentum dependent.
Hence, instead of attempting
to solve the AG equations for the general $\rm u_{\bf k k'}$ used above,
we restrict ourselves here to the case
$\rm u_0$ = 1, $\rm u_1$ = 0, i.e. only isotropic impurity scattering.
The results are shown in Fig. 2(c) and (d). 
It is obvious that in the limit of strong 
scattering the gap amplitude is reduced more rapidly by the
introduction of impurities than in the Born limit 
(comparing Fig. 2(a) and (c)).
When approaching the critical regime by decreasing $\rm V/V_c $,
a reduction of the gap amplitude (position of the peak in $\rm
N(\omega )$) and an increase in the residual DOS is observed,
similar to the Born limit treated above. However, since only isotropic
impurity scattering was considered in the latter case there is no
widening of the node regions in the gap function 
and hence no additional enhancement
of N(0).
While from the experimental side it has not been settled 
what the strength of the effective scattering potential,
${\rm |u_{\bf k,k'}|}$, should be,
a comparison of the two extreme limits suggests that no dramatic 
differences are to be expected within AG
calculations.\cite{borkowski} Qualitative differences between these 
two limits do appear for the single impurity problem, 
where anisotropic impurity resonances 
appear in the unitary limit.\cite{bsr} 

Let us now turn to the qualitative changes in the shape of the gap
function in the presence of a momentum dependent impurity scattering 
potential.
In the clean case, there are
four nodes located at $\phi = (2 n - 1) \pi/4$ (n=1,...,4)
(Fig. 1(a)). The
first of these nodes is shown in Figs. 3(a) and (b) for 
a fixed impurity concentration ($\rm \Gamma / \Gamma_c = 0.15$)
and various values of V. 
In Fig. 3(a) only isotropic impurity scattering was considered
($\rm u_0 = 1$, $\rm u_1 = 0$) while in Fig. 3(c) the case favoring
forward scattering is shown ($\rm u_0 = u_1 = 1$). In the latter 
case the gap function was extracted by analyzing the angle resolved
DOS: $\rm \Delta (\phi )$ was defined as the position of the 
inflection point in $\rm N(\phi ,\omega )$ (maximum
in $\rm \partial N(\phi , \omega )/\partial \omega $). In both cases the
gap amplitude decreases with decreasing V. (A similar behavior occurs 
when V is kept fixed and $\Gamma$ is increased.)
In the case of forward scattering, however, a flattening of the gap
function around the nodes is observed. To quantify
this behavior, we plot the extension, $\rm d_{\phi } $, of the ``flat" part
of $\Delta (\phi )$ in the insets. 
With $\rm d_{\phi } $ we denote those segments
of $\Delta (\phi )$ 
where the magnitude of $\Delta (\phi )$ has fallen below half of
its maximum value, $\rm d_{\phi } = \pi /2 - 2 \phi( \Delta_{max} /2 )$.
\cite{footnote3}
Note that as long as only a finite number of harmonics in the expansion
of $\rm \Delta (\phi )$ is considered, only extended saddle points can
occur. However, it is obvious from the above discussion that there is 
a clear tendency towards gapless (``normal") areas in the presence of
momentum dependence in the impurity scattering potential 
$coexisting$ with gapped regions (as depicted schematically in Fig. 1(b)). 
This prediction 
may be verified in overdoped cuprate superconductors
by tunneling experiments, and by
angle resolved photoemission
spectroscopy (ARPES) if an adequate energy resolution (of order 1 meV) can be
achieved.\cite{arpes}
 
\begin{figure}[hp]
\centerline{\psfig{figure=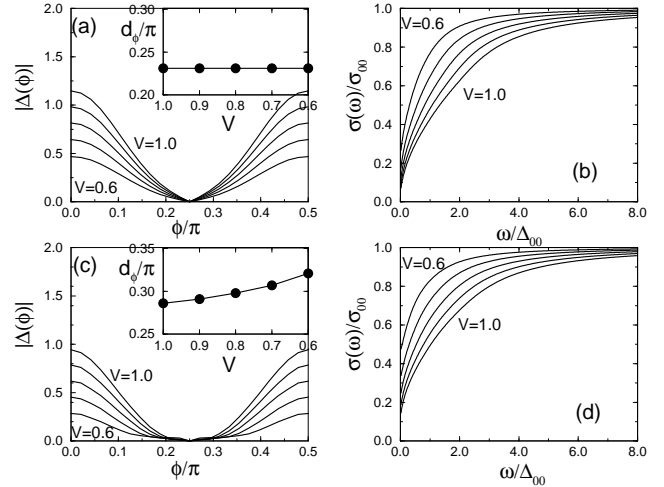,height=6.5cm,angle=-90}}
\vspace{0.5cm}
\caption{
(a) Angular dependence of the magnitude of the gap function in the Born limit 
at $\rm \Gamma / \Gamma_{0c} =0.15$, $\rm u_0 =1$, $\rm u_1 = 0$,
and $\alpha = 0.2$. The inset shows the length of the segment
of the gap function
where $|\Delta (\phi ) / \Delta (0)| \leq 0.5$. 
(b) Optical conductivity corresponding to the parameters chosen
in (a). $\sigma_{00}$ is the normal state value of the optical
conductivity.
(c) As (a), but with $\rm u_1 = 1$. 
(d) As (b), but with $\rm u_1 = 1$.
}
\end{figure}

In the critical regime, where the amplitude of the gap function is
small, Eqs. (3-5) can be linearized. While the renormalization of
$\Delta_0 $ remains frequency dependent in this limit, a functional
form of the renormalized gap function can be extracted from the angular
dependent residual DOS, $\rm N(\phi ,0)$. We find that 
$\rm \tilde{\Delta}_0 = \Delta_0 [1 + {u_1^2}/{(4 u_0^2 +2 u_1^2)} ]^{-1}$,
while $\Delta_1 $ remains unrenormalized. 
For the above choice of parameters ($\rm u_0 = u_1 =1$, $\alpha =0.2$),
we then find $\rm \tilde{\Delta}_1/\tilde{\Delta}_0 \approx$ 0.233.

A complementary experimental probe that is sensitive to extended flat regions
in the gap function is microwave spectroscopy, which measures the real
part of the c-axis infrared optical conductivity,
\bea
\rm
\sigma (\omega )=\frac{\pi e^2}{\omega }\int_0^{\omega}d \omega '
\int\frac{d^2 {\bf k}}{(2 \pi)^2}
tr[A_{\bf k}(\omega -\omega ')
A_{\bf k}(-\omega ')],
\eea
where $\rm A_{\bf k}(\omega ) = -\frac{1}{\pi} 
Im[{\hat{\tilde{G}}}_{\bf k}(i\tilde{\omega}_n) ]|_{i\tilde{\omega}_n =
\tilde{\omega} + i\delta} $,
and an angular average has been taken over the in-plane directions.
\cite{quinlan}
In Fig. 3(b) and (d), the optical conductivities are shown for the
two cases of isotropic and forward impurity scattering. 
In the clean case, $\sigma (\omega )$ shows a ``knee" feature at 
2$\rm \Delta_{00}$ ($\rm \Delta_{00} \equiv \Delta (\Gamma = 0, V =1.0)$). 
In the presence of impurities, this energy scale
is reduced when approaching the critical regime by decreasing V.
In analogy to the discussion of the DOS, the increase of 
$\sigma (0)$ is enhanced considerably in the case of forward scattering.

In Fig. 4(a) and (c) the residual DOS is shown for the Born limit
at $\rm u_0 = 1$, $\rm u_1 = 1$, and $\alpha = 0.2$, and for the 
unitary limit at $\rm u_0 = 1$, $\rm u_1 = 0$, and $\alpha = 0$.
By expanding about the clean limit, it is seen that 
in the Born limit, the zero-frequency DOS vanishes exponentially
slowly, $\rm N(0) \propto \exp(-const/\Gamma )$, for small impurity 
concentrations. On the other hand, in the unitary limit an infinitesimal 
amount of impurities is sufficient to yield a finite residual DOS,
$\rm N(0) \propto \sqrt{\Gamma } $. Hence an enhancement of the residual
DOS can be induced both by increasing the strength of the scattering 
potential and by introducing angular dependence in it. However, only
the latter case leads to extended flat regions in the gap function.

As a consequence of the finite residual DOS, the low-temperature
London
penetration depth has the form $\rm \lambda (\Gamma , T) = \lambda (0,0) +$ 
$\rm  N(\Gamma , 0)/2 +$ $O(\rm T^2)$.\cite{scalapino}
As the effective electron-electron interaction strength is decreased, 
i.e. when introducing additional holes into optimally doped or overdoped
cuprate superconductors, $\rm \lambda $ grows rapidly.
This behavior, illustrated in Figs. 4(b) and (d),
is in agreement with recent measurements by 
Locquet {\it et al}.\cite{locquet} 

\begin{figure}[hp]
\centerline{\psfig{figure=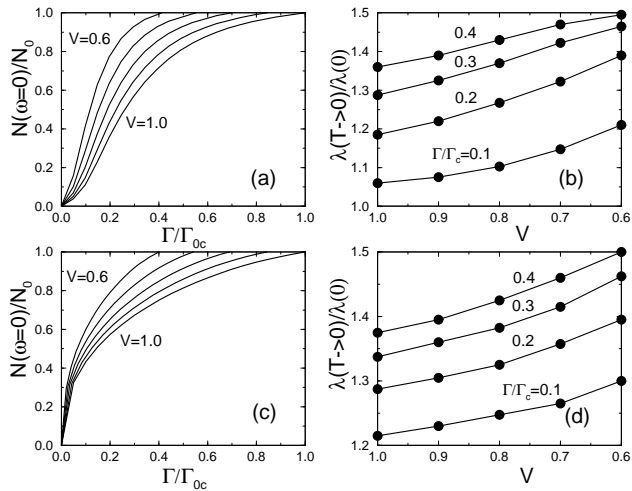,height=6.5cm,angle=-90}}
\vspace{0.5cm}
\caption{
(a) Residual DOS as a function of impurity concentration
in the Born limit. Here $\rm u_0 = 1$, $\rm u_1 = 1$, and $\alpha = 0.2$.
(b) Low-temperature London
penetration depth as a function of the $\rm e^--e^-$
pairing potential. The parameters are chosen as in (a).
(c) As in (a), but in the unitary limit, and with 
$\rm u_0 = 1$, $\rm u_1 = 0$, and $\alpha = 0$.
(d) Low-temperature London penetration depth for the parameters 
chosen in (c).
}
\end{figure}

In our results the maximum value of the gap is reduced consistently
as either $\Gamma $ is increased or V is reduced. This agrees with the
results of ARPES and tunneling experiments. It disagrees with the conclusion 
from studies of the optical conductivity. However, there is an increase in the
gapless region in the case of forward weighted impurity scattering which
goes some way towards explaining
the optical experiments.
Our conclusion is 
that variations in impurity character may be at least partly 
responsible for the divergence of the experimental results, although
an explanation for the absence of renormalization of the gap maximum
is not possible within this weak coupling scheme.

We wish to thank D. Agterberg,
D. Duffy, A. Nazarenko, B. Normand, and A. van Otterlo
for useful discussions,
and acknowledge
the Swiss National Science Foundation 
and the US Department of Energy for financial support.

%
%


\begin{references}
\bibitem{footnote1} 
Although in the zero-temperature limit the frequency sums are replaced
by integrals,
the indices `n' for the Matsubara frequencies,
$\rm \omega_n$, are kept
to remind the reader that the AG equations are written in imaginary
frequency space. For an evaluation of the corresponding
DOS an
analytic continuation onto the real frequency axis is required.
\bibitem{footnote2} In the absence of impurities ($\Gamma =0$), clearly
$\rm V_c =0$, corresponding to the usual BCS instability.
However, in the presence of disorder 
the critical coupling $\rm V_c(\Gamma )$ is finite.
\bibitem{footnote3} Here we have inverted $\Delta (\phi )$ to extract the
extended flat region of the gap function as defined by $\rm d_{\phi } $.
\end{references}
\end{document}